\documentclass[pre,twocolumn,aps]{revtex4}
\usepackage{graphicx}
\usepackage[activeacute,english]{babel}
\begin{document}
            
\title {
Symmetric Brownian Motor
}
\author{
A. Gomez-Marin$^{1}$
}
\author{
J. M. Sancho$^{1}$
}
\affiliation{
$^{1}$
Departament d'Estructura i Constituents de la Materia, 
Facultat de Fisica, Universitat de Barcelona, Diagonal 647, 08028 Barcelona, Spain}

\date{\today}

\begin{abstract}

In this paper we present a model of a symmetric Brownian motor (SBM) which 
changes the sign of its velocity when the temperature gradient is inverted. The velocity, external 
work and efficiency are studied as a function of the temperatures of the baths and 
other relevant parameters. The motor shows a current reversal when another
parameter (a phase shift) is varied. Analytical predictions and  results from 
numerical simulations are performed and agree very well. 
Generic properties of this type of motors are discussed.

\end{abstract}

\maketitle

\section{Introduction}

We all do know that it is possible to extract some amount of mechanical work 
from a thermal bath at a temperature $T_2$ provided we have another bath at a lower 
temperature $T_1 < T_2$. Thermal engines are the devices that perform this 
task. All this is well known from elementary textbooks on 
thermodynamics.  We also know from statistical mechanics that any object in a 
thermal bath exhibits random energy fluctuations of the order $K_BT$. These 
fluctuations are relatively very small for macroscopic objects but of very 
important relevance for nanometric objects such as biological motors: 
kinesins, dyneins, etc. \cite{Oster}. We are also familiar with windmills, 
which are able to 
extract useful work from random winds by a proper adaptation to the wind 
direction.  We can 
ask ourselves if it is possible to rectify thermal fluctuations by some 
appropriate mechanical devices. 

The engines which aim to get useful work by rectifying thermal 
fluctuations are called Brownian motors (BM). 
In fact, the paradigm of such speculations is Feynman's famous ratchet and
pawl machine \cite{feyn}. 
During the last years a lot of effort has been invested  to
study what has been called as the ratchet effect. This is a  
mechanism which consists in breaking the spatial and temporal inversion symmetry of the
system so that directed transport emerges, often
enhanced by the thermal fluctuations. 
The ratchet mechanism can be implemented in different ways. 
Here we will make a model through an equation for  a 
dynamical classical variable (position 
or angle) moving in a
periodic and asymmetric potential (a ratchet potential) coupled with 
another degree of freedom which will break thermal equilibrium.

\begin{figure}
\begin{center}
 \includegraphics[ width=0.35\textwidth]{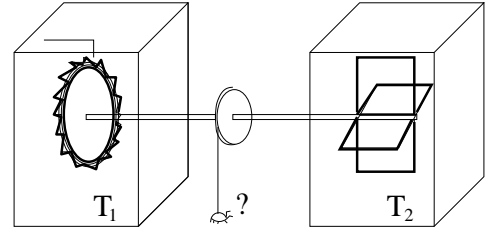}
   \caption{Feynman's original ratchet and pawl machine.}
   \label{frp}
\end{center}
\end{figure}

\begin{figure}
\begin{center}
 \includegraphics[ width=0.32\textwidth]{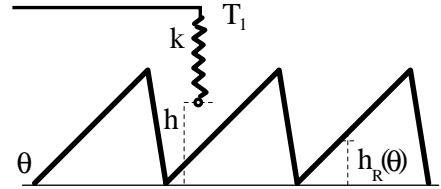}
   \caption{Detail of the stopping mechanism based on the
interaction between the ratchet and the pawl.}
   \label{dfrp}
\end{center}
\end{figure}

There is an enormous variety of ratchets in the literature:  pulsating
ratchets (on-off, fluctuating potential, traveling potential), tilting
ratchets (fluctuating force, rocking ratchets), Seebeck ratchets,
 Feynman ratchets, temperature
ratchets, frictional ratchets, quantum ratchets, collective ratchets,
mechanico-chemical ratchets, ionic bombs and pumps, etc. \cite{reiBM}. 

In this work we will 
focus on  a new Brownian motor inspired by the so called Feynman's ratchet 
(Fig.\ref{frp}). 
Feynman stated in his famous lectures \cite{feyn} that
a very particular  device (the ratchet and  pawl 
machine) could 
have the efficiency of a 
Carnot engine  when operating reversibly. 
He  idealized a gadget (see Fig.\ref{frp}) made up of two boxes 
with some gas at different
temperatures. The hotter box contains an axle with vanes in it. The 
bombardments of gas
molecules on the vane make the axle rotate with random symmetric fluctuations. 
At the other end side of
the axle there is  a second box with a toothed wheel which in principle 
can turn only one way. The pawl (the stopping mechanism) is under 
the influence of another
temperature (Fig.\ref{dfrp}). At first glance one could think that it seems quite likely
that the wheel will spin round one way and lift a weight even
when both gases are at the same temperature and thus violating the Second Law. However, 
a closer look at the pawl reveals that it bounces and so the wheel will rotate randomly in any 
direction, doing a lot of jiggling and with no net turning. Thus, the machine
cannot extract work from two baths at the same temperature.

When the temperature of the vanes is
higher than the temperature of the wheel, Feynman concludes that
some work is performed with Carnot's
efficiency when the machine is lifting the weight very slowly. 
This is indeed a very optimistic
result which has been revised,
many years later, in Refs. \cite{parr,magfrp}. Indeed, 
there are some overlooked aspects in
Feynman's argument which can be summarized as follows: since the engine seems
to be simultaneously in contact with two
baths at different temperatures (through the rigid axle), it cannot work in 
a reversible way and Carnot's efficiency will never be achieved. 
We will come back to this point along this paper.

A possible mathematical model of the device of Fig.\ref{frp} and Fig.\ref{dfrp} in 
terms of overdamped Langevin equations is, 
\begin{equation}
\lambda_{1} \frac{d h}{dt}= -\partial_{h} V+\xi_{1}(t),
\end{equation}
\begin{equation}
\lambda_{2} \frac{d \theta}{dt}= -\partial_{\theta} V+\xi_{2}(t),
\end{equation}
where $\xi_i(t)$ mimics thermal fluctuations and it is assumed to be a white 
noise, Gaussian distributed with
zero mean,  satisfying the
fluctuation-dissipation theorem,
\begin{equation}
<\xi_i(t)\xi_j(t')>=2k_{B}T_{i}\lambda_{i} \delta(t-t') \delta_{ij}.
\label{noisecor}
\end{equation}

The total potential energy  is modeled by,
\begin{equation}
V(\theta,
h)= \tau \theta + \frac{1}{2}kh^{2} + \frac{e}{e^{\frac{|h-h_{R}(\theta)|}{l_{0}}}-1},
\end{equation}
where, $h_{R}(\theta)$ represents the periodic but asymmetric profile of the
ratchet. An explicit expression will be given in the next section.

The potential $V$ has three terms.
$\tau$  is the torque which gives useful work. The second term
accounts for the potential energy stored in the spring (the pawl) with 
constant $k$, which pushes down the ratchet.
Finally, the last term is
a very repulsive potential at short distances and nearly zero at long ones. It
is used to avoid that the pawl crosses through the real physical surface of
the ratchet. At the same time, it couples both degrees of freedom $\theta$ and
$h$. 
Other potentials can
account for the same physics in this mathematical scheme \cite{magfrp,seki}.

Numerical simulations show a non-zero mean velocity of the ratchet 
device for very high temperature
differences. The  main conclusion is that, although the motor can perform
useful work, it has an extremely small efficiency, 
being very far away from Carnot's efficiency 
estimated by Feynman, as pointed out and shown in
Refs. \cite{parr,magfrp,seki,tsuo}. This motor presents a high thermal  
conductivity and as a consequence has a very low efficiency. 

The structure of this paper is the following. In section II we present the
model for a symmetric Brownian motor and the 
numerical results obtained by computer simulations of the equations of such device.
Section III is devoted to the analytical approach to this model. We end 
in Section IV with some comments and conclusions.

\section{A symmetric Brownian motor (SBM)}

Feynman's ratchet, and similar models,   do not 
fulfill the following inversion property: 
$T_2 \leftrightarrow T_1  \, \longrightarrow  v \leftrightarrow -v$.   
Thus our initial motivation to propose this new model is to have a motor 
such that, when switching the temperatures of the two baths, 
this is, when reversing the
temperature gradient, the mean velocity absolute 
value does not change while the sign of the velocity does. 
Obviously, this would imply  some  geometrical symmetries in the engine.  
One example of a microscopic model for a Brownian motor
 that suffers this change in the sign of the
current can be found in \cite{van}.  There is also a model for a diode
rectifier that is analytically solvable  \cite{igor} and that has the same 
main features as the SBM,
namely, it is symmetric and it shows "reversibility".

\begin{figure} [t] 
\begin{center}
  \includegraphics[ width=0.4\textwidth]{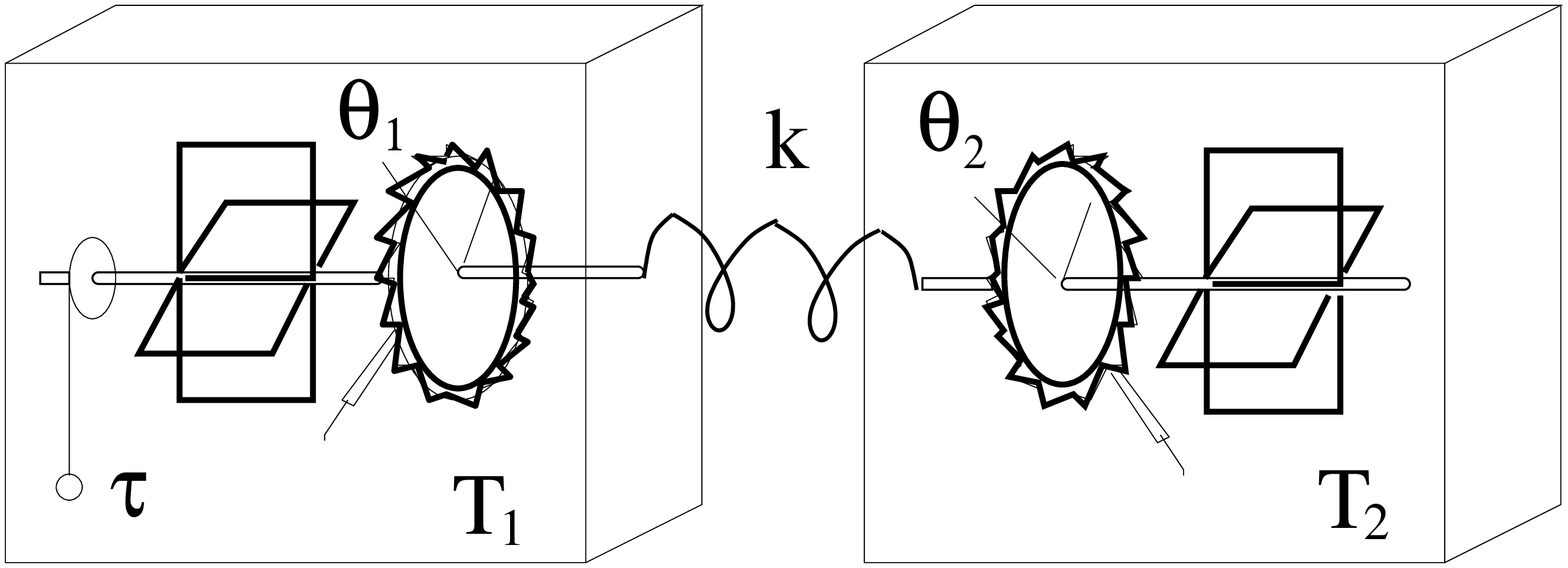}
  \caption{Model for the symmetric Brownian motor (SBM).}
  \label{motorsym}
\end{center}
\end{figure}

Our "symmetric" motor is shown in Fig.\ref{motorsym}. 
According to this scheme, the stochastic differential equations that 
describe the dynamic evolution of the system in the over-damped regime are,
\begin{eqnarray}
\lambda_{1} \frac{d \theta_{1}}{dt}= -\partial_{\theta_{1}} V+\xi_{1}(t),
\nonumber\\
\lambda_{2} \frac{d \theta_{2}}{dt}= -\partial_{\theta_{2}} V+\xi_{2}(t).
\label{model3}
\end{eqnarray}
The noise terms are again white thermal noises, Gaussian distributed and with
zero mean, satisfying the fluctuation-dissipation theorem (\ref{noisecor}).

The potential $V$,
 contains an external torque $\tau$, two ratchet-shape potentials
and an harmonic interaction that
 couples both degrees of freedom,
\begin{equation}
V(\theta_{1},\theta_{2})=\tau \theta_{1}+ V_{R}(\theta_{1})+V_{R}(-\theta_{2})+
\frac{1}{2}k(\theta_{1}+\phi-\theta_{2})^{2},
\label{potential3}
\end{equation}
where the phase shift $\phi$ is one of the most important parameters.

\begin{figure} [t]
\begin{center}
\includegraphics[angle=270, width=0.35\textwidth]{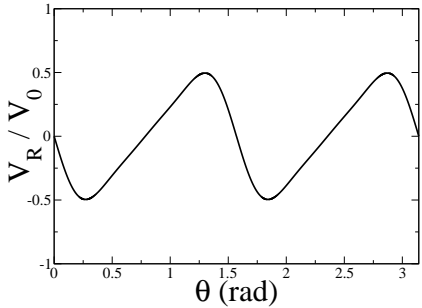}
\caption{Shape of the ratchet potential $V_{R}$ for $d=4$.}
\label{rat}
\end{center}
\end{figure}

In Fig.\ref{rat} we see the form of the ratchet potential $V_{R}(\theta)$ used in the
numerical simulations. Its analytical expression is,
\begin{equation} \label{pot}
V_{R}(\theta) = -\frac{V_{0}}{2.23} \mathcal{V}(\theta),
\end{equation}
\begin{equation}
\mathcal{V}(\theta) = \sin(d \theta)+0.275\sin(2d\theta)+0.0533\sin(3d\theta).
\end{equation}
$V_{0}$ controls the height of the
potential,  $d$ is the number of teeth per cycle $2\pi$, and
the asymmetry of the potential is controlled  by changing the numerical
coefficients that multiply the
sinus functions expansion of the potential.

 Notice that the ratchet
potential that one variable $(\theta_{1})$ sees is the specular image of the one that the
other variable $(\theta_{2})$ feels, as in the diode rectifier of Ref. \cite{igor}. It is 
in this sense that we call our model
symmetric.

To study the relevance of each parameter we 
proceed to express our system in terms of dimensionless ones.
Introducing the dimensionless time $s$ as 
 $t=\frac{\lambda_{1}}{V_{0}}s$,
the set of equations (\ref{model3}) becomes more compact. 
The dimensionless parameters are now: 
$\widetilde{T_{1}}=\frac{k_{B}T_{1}}{V_{0}}$,
$\widetilde{T_{2}}=\frac{K_{B}T_{2}}{V_{0}}$, $\widetilde{k}=\frac{k}{V_{0}}$,
$\widetilde{\lambda}=\frac{\lambda_{1}}{\lambda_{2}}$,
$\widetilde{\tau}=\frac{\tau}{V_{0}}$ and, of course, $\phi$, which is in
radians. Notice that $V_{0}$ controls
the energy scale, which is measured in $K_{B}T_{1}$ units. We 
also see, for
instance, that only the fraction of the friction coefficients is relevant.

\subsection{Numerical results of SBM}

\begin{figure}
\begin{center}
  \includegraphics[  angle=270,  width=0.4\textwidth]{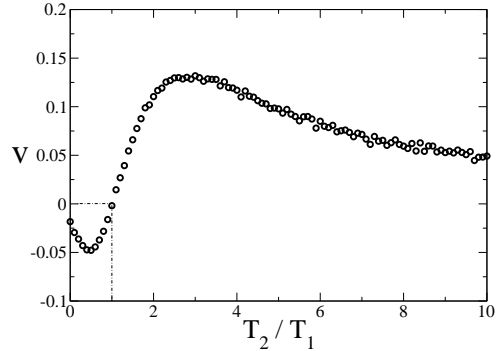}
  \caption{Mean velocity versus $T_{2}/T_{1}$ ($\tau=0$).}
   \label{T2}
  \end{center}
\end{figure}

Preliminary numerical results indicate that the motor inverts its velocity when 
the temperature gradient is also inverted (see Fig.\ref{T2}). Some 
important thermodynamic requirements are also fulfilled: there is no net 
motion in the 
limits $T_2 \rightarrow T_1$ (a unique bath), $k \to
0$ (no coupling between baths) and $k \to \infty$ (only one effective 
temperature).

\begin{figure}  
\begin{center}
  \includegraphics[  angle=270, width=0.4\textwidth]{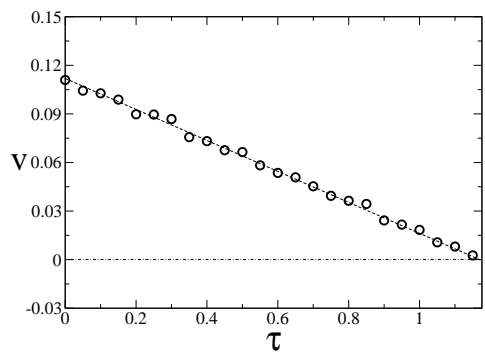}
  \caption{Mean velocity of the motor at the optimal regime as a function of
    the external torque $\tau$.  
The stall torque (where $v=0$) is around $\tau=1.2$.}
   \label{vtau}
\end{center}
\end{figure}

To perform the simulations we take $\widetilde{\lambda}=1$ for simplicity and $K_{B}T_{1}=1$. 
As in previous models for motors we have explored the parameter domain to get the most 
effective values (larger velocities). This situation corresponds to the 
parameter values $V_{0}=2.5$, $k=100$ and $d=16$. In Fig.\ref{T2} and Fig.\ref{vtau} some numerical 
results of the mean velocity $v \doteq \langle \dot{\theta_{1}} \rangle $
are presented for $\phi=0.4$. We see in Fig.\ref{T2} that there is a  
maximum around $T_{2}/T_1=2.5$ and a minimum at $T_{2}/T_1=0.5$. This implies
that the velocity is bounded and larger temperature gradients do not imply larger 
velocities. 
For very large $T_2$ the motor does not see the ratchet potential and cannot
take advantage of the broken inversion symmetry. Then the motor has a zero
mean velocity.
The linear dependence of the velocity around  $T_2/T_1 \sim 1$ is is clear 
signature that the motor has the inversion property. We will
extend this result further. In Fig.\ref{vtau} we see the expected negative linear dependence of the velocity 
versus torque until the stall force where the velocity is zero.  
  
\begin{figure}  
\begin{center}
  \includegraphics[  angle=270,  width=0.4\textwidth]{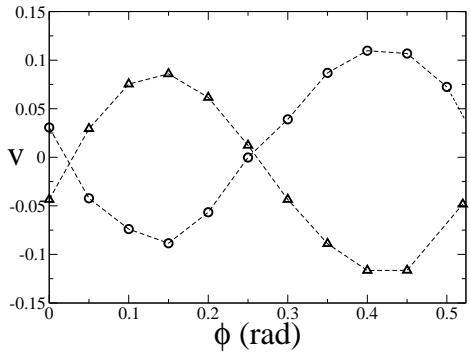}
  \caption{Current inversion profile $v(\phi)$ when  $T_{2}=2T_{1}$ (circles)
    and when $T_{1}=2T_{2}$ (triangles).  
The angle goes from $0$ to $0.52$ because we are
 considering $d=12$ and therefore the periodicity is  $\frac{2 
\pi}{12}=0.52$. The dashed lines are a guide to the eyes.} 
  \label{ci}
\end{center}
\end{figure}

Moreover the most striking result is the velocity inversion as a 
function of the phase difference parameter $\phi$.
We show in Fig.\ref{ci} the mean velocity of the motor as a function of the phase shift
$\phi$ in the cases of $T_{2}=2T_{1}$  and when $T_{1}=2T_{2}$. 
One must notice that both cases are "symmetric".  This result allows to extract 
some conclusions: the existence of a current inversion as a function of 
$\phi$, the great
amplification of the speed for the particular value of $\phi=0.4$  
and, finally, that the motor is  perfectly symmetric when
reversing the temperature gradient. 

These results deserve theoretical explanations that will be presented in
section III.

\subsection{Energetics of the SBM: the efficiency}

It is not enough to know whether a motor will move, nor even how fast it
will run. One would also like to know how efficiently such machine can operate. 
There are many different approaches in the literature to study
energetics and, in particular, efficiencies \cite{seki,suzu,lucz}.  
We will follow  Sekimoto's
characterization of the energetics of thermal ratchet motors \cite{seki}. 
This 
scenario is the most suitable scheme both because it is
intuitive and also from its potential applicability.  The main idea is to find
how much energy, received from an external source, the 
motor will be able to employ to produce useful mechanical work, i.e. to lift a
load. In order to compute this, we must understand where the energy
comes from, where it can go and what ingredients account for that.
Sekimoto deals with energies and not
with power. When one tries to study the problem in
units of power, velocities appear explicitely in the computations and,
since they are instantaneously not continuous because of the stochasticity of
the system, one finds that some of the terms are extremely bad behaved and it
is impossible to get a reasonable numerical convergence in the
simulations. Moreover, we 
have found it easier to directly compute the stochastic integrals and
afterwards perform the average. 

Let's analyze the energetics of such kind of systems. We would expect there's
some energy $R$ that is released from the bath at $T_2$ (the one at a higher temperature)  
and that is partially converted into mechanical work $W$ to lift the load. 
However, according to the Second Law,  some of
this input energy must be dissipated $D$ into the heat bath at $T_1$. If the 
system has
an intrinsic potential there will also be a change in the potential energy
denoted by $U$. So the law of the total energy
conservation is,
\begin{equation} 
\label{balance}
R=D+W+U.
\end{equation}

Our interest here is to find the efficiency of the motor, which is defined, 
in general, as ratio between the work extracted versus the energy input,
\begin{equation}
\eta=\frac{ \langle W \rangle }{\langle R \rangle }.
\end{equation}
The work $W$ is quite easy to calculate since it is just the torque 
times the angle displaced. Then 
the work performed during a time interval $t_{f}-t_{i}$ can be obtained as,
\begin{equation}
\langle W \rangle= \tau \langle \theta_{1}(t_{f})-\theta_{1}(t_{i}) \rangle =
\tau \langle v \rangle (t_{f}-t_{i}).
\label{work}
\end{equation}

\begin{figure} 
\begin{center} 
  \includegraphics[ angle=270, width=0.4\textwidth]{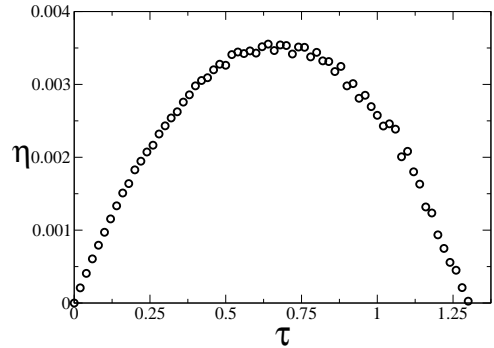}
   \caption{Efficiency $\eta$ as a function of the torque $\tau$. 
The numerics are done at the optimal regime of the SBM.}
   \label{eff}
\end{center}
\end{figure}

The evaluation of $ \langle R \rangle $ 
 needs a much more careful analysis. Following  \cite{seki}, we can write, 
\begin{equation} 
\label{R}
R=\int_{t_{i}}^{t_{f}} [\frac{\partial V(\theta_{1}(t), 
\theta_{2}(t))}{\partial \theta_{2}}]d\theta_{2}(t).
\end{equation}
So by using  (\ref{potential3}), we find that the energy that the bath at 
$T_2$ transfers to the mechanism in the bath at $T_1$ is, 
\begin{equation} 
R=\int_{t_{i}}^{t_{f}} [-k(\theta_{1}(t)+\phi-\theta_{2}(t))+ 
V_{R}'(-\theta_{2}(t))]d\theta_{2}(t).
\end{equation}
We have to evaluate this integral for different realizations and perform 
usual statistical averages. We have observed that this quantity is always 
large thus implying important looses of heat. 

To progress further let us first note that the potential 
energy $U$ does not contribute because we have a cyclic engine. In fact  
$V_{2}'(-\theta_{2}(t))$  is periodic and 
bounded and thus it can be discarded in the
long time interval studied.

Thus the final expression for the efficiency that has to be evaluated 
numerically is,
\begin{equation}
\eta=\frac{\tau \langle  \theta_{1}(t_{f})-\theta_{1}(t_{i}) \rangle }{-k \langle 
\int_{t_{i}}^{t_{f}} (\theta_{1}(t)+\phi-\theta_{2}(t))d\theta_{2}(t) \rangle}.
\end{equation}
In Fig. \ref{eff}, the efficiency $\eta$ as a function of the torque $\tau$ 
 is plotted.
The specific values we use are around those that were shown to be the optimal
ones; $V_{0}=3$, $T_{2}=2$, $k=100$, $\phi=0.4$ and 
$d=12$. The plot shows a parabolic curve for the efficiency with a maximum 
at the middle value of the stall torque. Notice that the maximum of the  
efficiency is still extremely low. Let's also underline that at the stall
force ($\tau \simeq 1.25$), the
efficiency is zero, which means that unavoidable heat transfer occurs
\cite{parr}. Therefore, even when
$\langle v \rangle \to 0$ there's a non-zero heat flux transfered from the hotter bath to the 
cooler one. This is the reason why these motors have so low efficiencies.

\section{Analytical study of the current inversion}

As we have seen in the numerical results the $\phi$ parameter controls a 
spectacular phenomena. The SBM exhibits
velocity inversion when this parameter is varied. 
We know that this phenomenon
exists and has been studied in the literature \cite{rei}. Our purpose 
now is to predict analytically that this happens in our model. Any kind of
exact calculation in this system seems impossible and, so, some  
approximations have to be assumed.

Consider the equations that define our Brownian motor (\ref{model3}) written 
in the form,
\begin{equation}
\dot{\theta_{1}}=f(\theta_{1})-k(\theta_{1}+\phi-\theta_{2}) +\xi_{1}(t),
\end{equation}
\begin{equation}
\dot{\theta_{2}}= -f(\theta_{2})+k(\theta_{1}+\phi-\theta_{2}) +\xi_{2}(t),
\end{equation}
where $f(\theta)$ is the force exerted by the ratchet potential and there is
no torque. Let us define now the change of variables \cite{klu},
\begin{equation}
x=\frac{\theta_{1}+\theta_{2}}{2},
\end{equation}
\begin{equation}
y=\frac{\theta_{1}-\theta_{2}+\phi}{2}.
\end{equation}
The relevant variable $x$ describes the evolution of the center of mass
and the "irrelevant" variable $y$  describes the relative motion of   
the two-particle system. Introducing a redefinition of the noises,
\begin{equation}
\eta_{1}(t)=\frac{\xi_{1}(t)+\xi_{2}(t)}{2},
\end{equation}
\begin{equation}
\eta_{2}(t)=\frac{\xi_{1}(t)-\xi_{2}(t)}{2},
\end{equation}
and using the fact that $y$ is very
small, we can make a Taylor expansion up to first order, getting the  pair of equations,
\begin{equation}
\dot{x}=F(x)+yG(x)+\eta_{1}(t),
\label{xeq}
\end{equation}
\begin{equation} 
\label{yeq}
\dot{y}= Q(x)+ y (R(x)-2k)+\eta_{2}(t).
\end{equation}
The explicit expressions for these quantities and further mathematical details
are in the Appendix. Summarizing, we have found that the mean angular
velocity $v$ can be expressed in the standard form \cite{rei,risk,lutz},
\begin{equation}
v_{\phi}=\frac{2\pi}{d} N(1-e^{\beta_{\phi}}),
\label{vtheo}
\end{equation}
where $N$ is a kind of normalization constant that does not depend very much 
on the parameter $\phi$. The relevant quantity $\beta$, which is a function of
$\phi$, is given by,
\begin{equation} 
\label{beta}
\beta_{\phi}=\int_{0}^{L} dx \frac{-F(x)-\frac{Q(x)G(x)}{2k}} {\frac{1}{4} \left[
      1+\frac{T_{2}}{T_{1}}+(1-\frac{T_{2}}{T_{1}})\frac{G(x)}{k} +
O(1/k^{2})  \right]   }.
\end{equation}
This expression can be numerically evaluated using the expressions for the 
different functions that appear in it. The value of the parameter $N$ has 
been estimated by 
 using only one numerical value of the velocity obtained by the simulation (the 
minimum value at $\phi=0.25$). 
If $\beta$  changes its 
sign 
as a function of $\phi$ then the inversion phenomenon is predicted. 

\begin{figure} [t]
\begin{center}
  \includegraphics[ angle=270, width=0.4\textwidth]{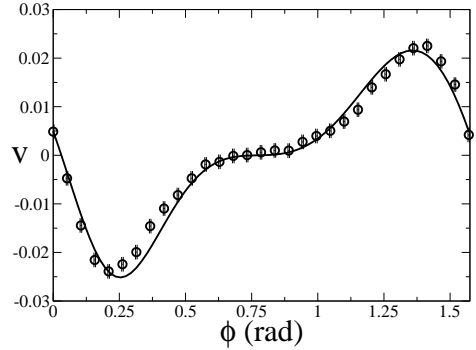}
  \caption{Comparison of $v$ vs $\phi$ obtained from numerical simulations 
of the model (\protect\ref{model3}) (circles) and from the 
analytical expression (\protect\ref{vtheo}) (solid line).
The  parameters chosen are $d=4$, $k=100$, $V_{0}=3$ and $T_{2}=2T_{1}$. 
}
\label{Vformal}
\end{center}
\end{figure}

In Fig.\ref{Vformal} we compare the predicted theoretical dependence of 
the velocity as a function of the phase $\phi$ to the values obtained by 
numerical simulations of the model. The analytical result, 
considering the amount of approximations involved, traces very accurately the 
current inversion phenomenon. The fit is very encouraging and enlighting.
The positions of the maxima are quite well
determined, the inversion of current is clearly coincident with the
simulations and even the little shift near $\phi=0$ and $\phi=\frac{2\pi}{d}$
is faithfully reproduced. This means that in our approximations we have kept 
the most dominant ingredients.
Note that the x-axis goes from $0$ to $1.57$ because we have imposed
$d=4$ ($\frac{2 \pi}{4}=1.57$) in order to safely use the analytical expressions derived in the Appendix.
However, for $d=12$ the computation of $\beta$ predicts qualitatively the current inversion too.

Now we will give an intuitive explanation of why the motor runs
either forward or backwards, depending on the value of the parameter
$\phi$. The original problem can be mapped into a
multiplicative noise scenario (see the Appendix for details) well documented in the literature
\cite{rei,lin,buti1,buti2}. Then we can straight forward identify in our case the
effective potential $V_{eff}(x)$ (\ref{Veff}), which does not need to be
asymmetric anymore, and the effective modulation
$g_{eff}(x)$ (\ref{geff}), which is responsible for the breaking of the
symmetries that allow a non-zero mean velocity in the motor. What is more, by
plotting them we can very intuitively see whether and why there is a forward
or backward flux \cite{lin,buti1,buti2}.

\begin{figure}
\begin{center}
  \includegraphics[ angle=270, width=0.4\textwidth]{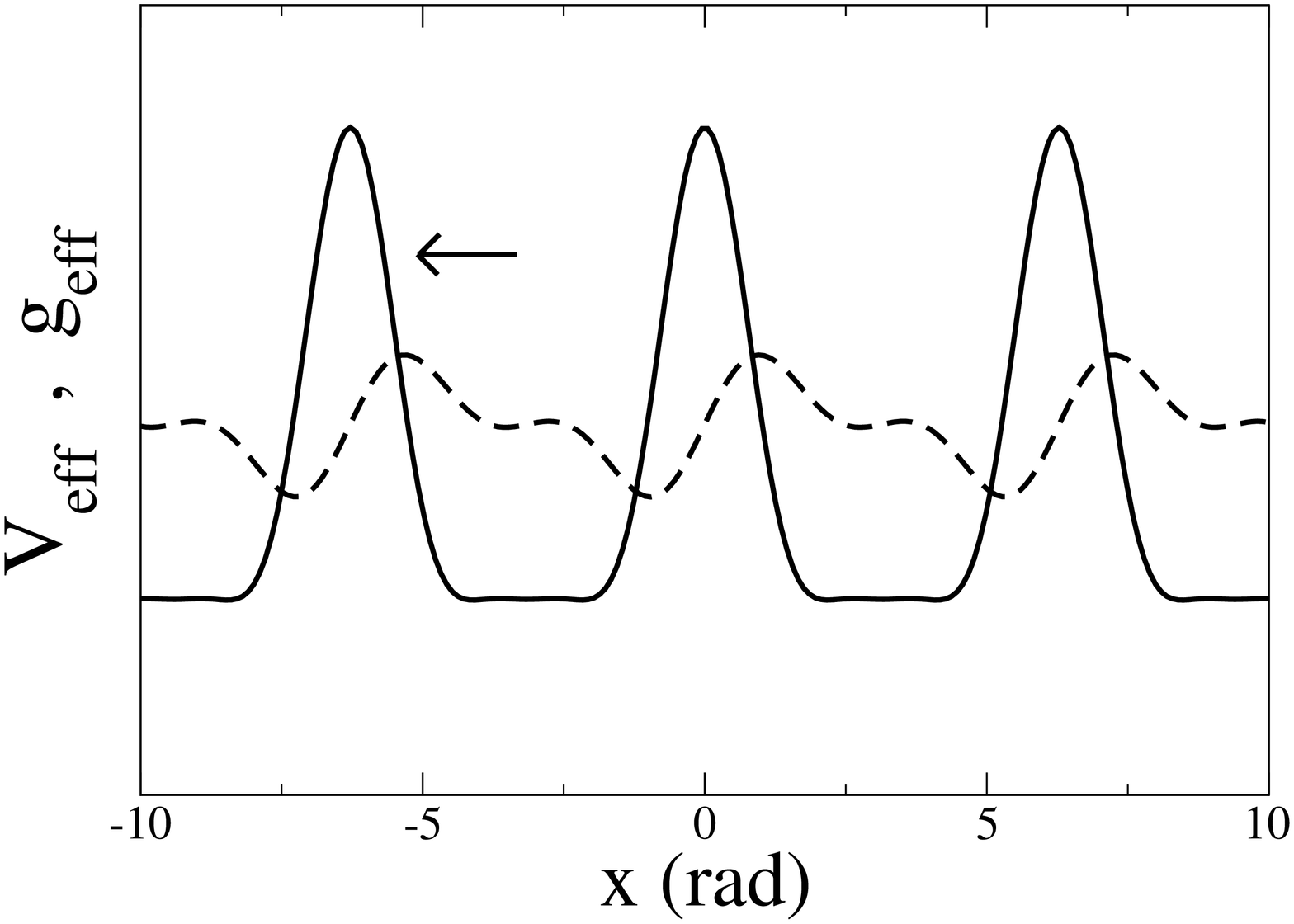}
  \includegraphics[ angle=270, width=0.4\textwidth]{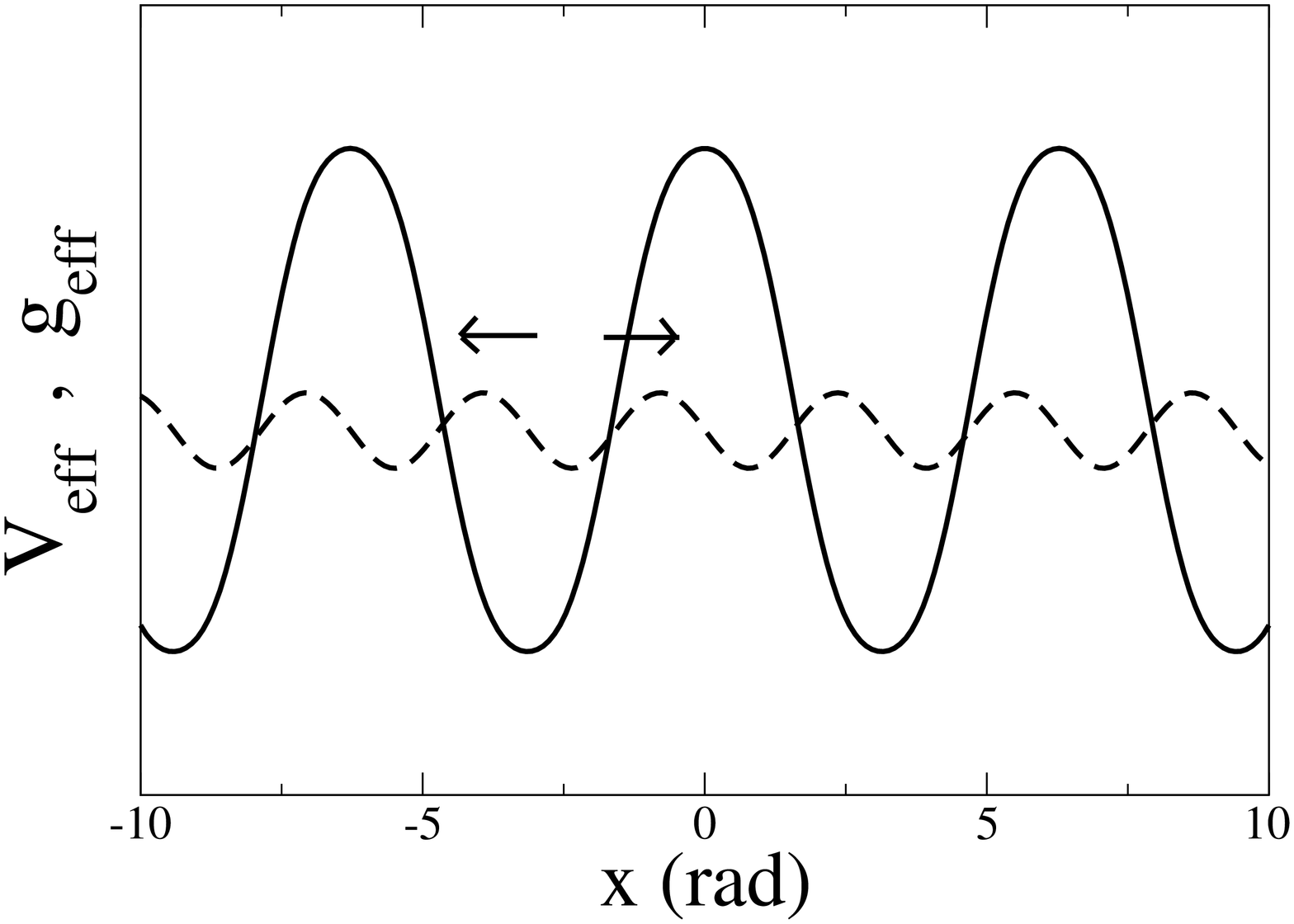}
  \includegraphics[ angle=270, width=0.4\textwidth]{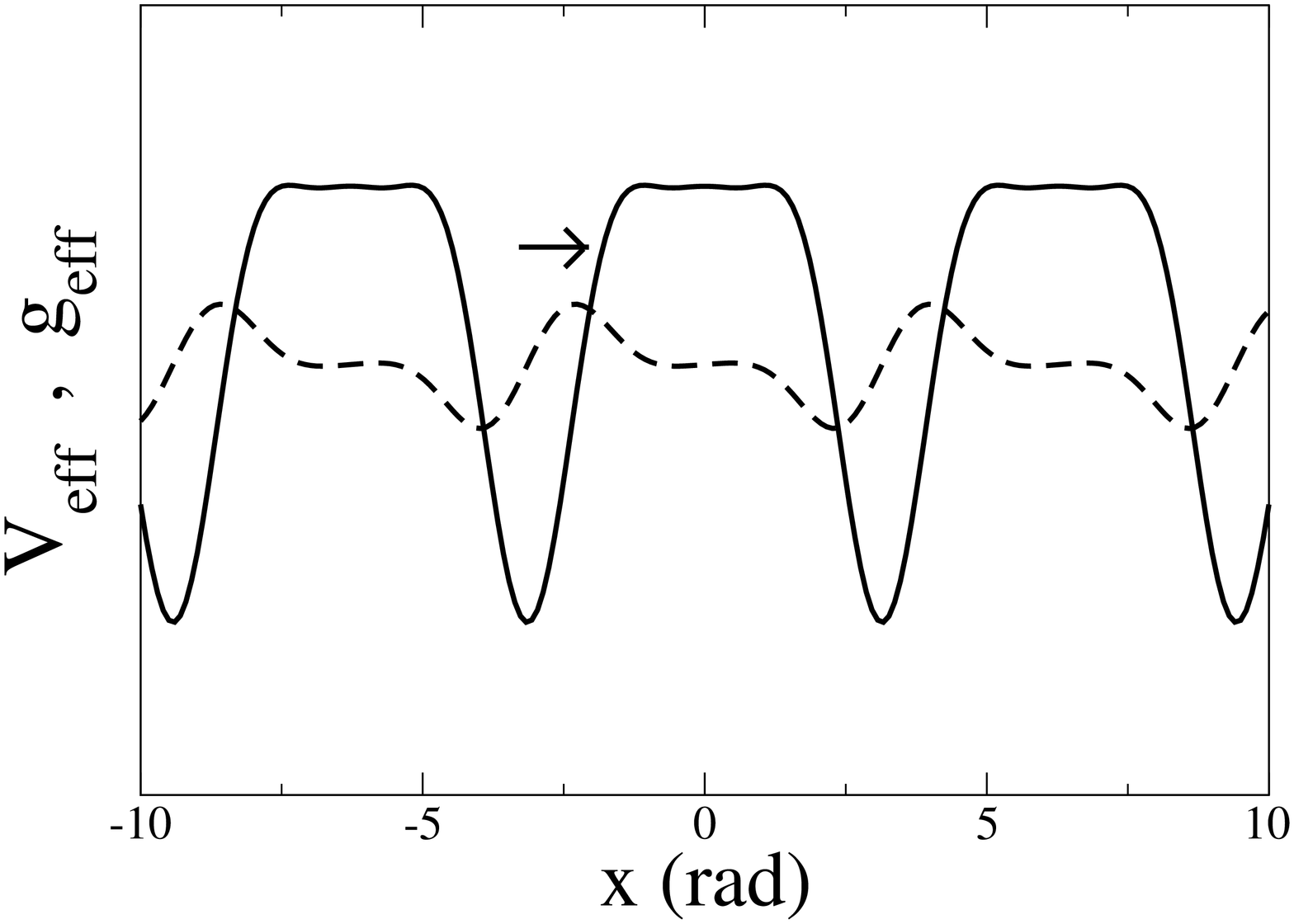}
  \caption{Plots of the shape of the effective multiplicative noise
  $g_{eff}(x)$ (\ref{geff}) (dashed line) and the effective potential
  $V_{eff}(x)$ (\ref{Veff}) (solid line) in arbitrary scales. The phase shifts are (from top to   bottom) $\phi=0.25$, $\phi=0.785$ and $\phi=1.375$, which correspond  to the fastest negative, zero and fastest positive velocities respectively for $d=4$. Arrows indicate the velocity direction.
}
\label{intu}
\end{center}
\end{figure}

In Fig.\ref{intu} we plot, for
three significative phase shifts, the noise coupling function $g_{eff}(x)$ and 
the
effective potential $V_{eff}(x)$. A careful look at each of them easily
reveals the current sign in a very intuitive way.
The key point is the fact that when $g_{eff}(x)$ is large, the noise effects are amplified  on one side of  
the hill of the periodic potential and, then, the net current will go in that
direction. Then, it is much clearer that depending on the position  of $g_{eff}(x)$
with respect to $V_{eff}(x)$ the current will go forwards, backwards or, in the
symmetric case, it will cancel. Thus, this simple scheme  explains the
qualitative behavior of the  current inversion. 

Two more comments are to be made.
Just for the sake of consistency, we have checked two basic conditions that must
always hold on these kind of motors. The first one concerns 
temperatures; if we set $T_{1}=T_{2}$, the integral $\beta_{\phi}$ vanishes because
the denominator becomes constant and, therefore, the average velocity is zero.
The second test is also related with the first one. When we make $k$ tend to
infinity (this implies a rigid coupling and
then only one effective temperature), the dependence on $x$ of
the multiplicative noise disappears again (only additive noise is present) 
and, then, the motor cannot move anywhere on average.

\section{Comments and conclusions}

We have presented and studied a symmetric Brownian motor. Its efficiency
has been numerically obtained and other specific properties have been studied.
This motor has a relevant external parameter (a phase) which induces
the phenomenon of current inversion.
We have performed an analytical calculation with appropriate approximations 
to
get an expression for the mean velocity in terms of the relevant
parameters of the model. This formal prediction fits very well the data from
the numerical simulations of SBM model.

An important conclusion is that, albeit the symmetric motor has a larger efficiency than
other mechanical Brownian motors, the efficiency of such devices is very
small. Regardless of the particular properties of these kind of heat
engines, they are anyhow unrealistic models for molecular motors since it is
known that these biological systems do transform chemical energy into work,
without the intermediate state of burning fuel. Consequently, one cannot think
of these models as realistic ones for biological molecular motors.

Moreover, the mechanical coupling mechanism between both baths acts as a very good heat conductor even in 
situations of very small mean velocity. Therefore, the efficiency is 
only a small fraction of that of Carnot. This is in fact a general feature of
heat Brownian motors due to the fact that in order to rectify thermal
fluctuations these systems must be put in contact and a lot of heat is interchanged.
The diode rectifier of Ref. \cite{igor} does not present this problem and 
accordingly it can arrive near the  Carnot efficiency in some limits.

\section{Acknowledgments}
Fruitful discussions with Professors I. Sokolov, J.M.R. Parrondo and C. Van
den Broeck are gratefully acknowledged.
This research was  
supported by the Ministerio de
Educaci\'on y Ciencia
(Spain) under project BFM2003-07850-C03-01.

\appendix
\section{Analytical approach}

Our starting point is the pair of coupled equations (\ref{xeq}) and (\ref{yeq}) 
where the different functions are,
\begin{equation}
F(x)=\frac{f(x-\phi/2)-f(x+\phi/2)}{2},
\end{equation}
\begin{equation}
G(x)=\frac{f'(x-\phi/2)+f'(x+\phi/2)}{2},
\end{equation}
\begin{equation}
Q(x)=\frac{f(x-\phi/2)+f(x+\phi/2)}{2},
\end{equation}.
\begin{equation}
R(x)=\frac{f'(x-\phi/2)-f'(x+\phi/2)}{2}.
\end{equation}

Since the variable $y$ has a faster dynamics we will first eliminate 
adiabatically equation (\ref{yeq}), and we will also discard the function $R(x)$ because the parameter $k$ 
is very large. Then equation (\ref{yeq}) reduces to,
\begin{equation}
y = \frac{1}{2k}\eta_{2}(t)+\frac{1}{2k}Q(x)
\end{equation}
Substituting now this expression in equation (\ref{xeq}) we get a Langevin equation 
with two multiplicative noises,
\begin{equation}
\label{eqx2}
\dot{x} =H(x)+g_{1}(x)\xi_{1}(t)+g_{2}(x)\xi_{2}(t),
\end{equation}
where the new functions are,
\begin{equation}
H(x)=F(x)+\frac{1}{2k}Q(x)G(x),
\end{equation}
\begin{equation}
g_{1}(x)=\frac{1}{2} \left( 1+\frac{G(x)}{2k} \right),
\end{equation}
\begin{equation}
g_{2}(x)=\frac{1}{2} \left( 1 -\frac{G(x)}{2k} \right).
\end{equation}

Let's write now the
Fokker-Planck equation associated to Eq. (\ref{eqx2}),
\begin{equation}
\partial_{t}P(x,t)=-\partial_{x}J(x,t).
\end{equation}
where,
\begin{eqnarray}
&  J(x,t)=H(x)P(x,t)-K_{B}T_{1} [g_{1}(x)\partial_{x}g_{1}(x)P(x,t)] &
\nonumber \\
& -K_{B}T_{2} [g_{2}(x)\partial_{x}g_{2}(x)P(x,t)]. &
\end{eqnarray}
After some manipulations
with partial derivatives, the probability current above can be rewritten as,
\begin{equation} \label{flux}
J(x,t)=H(x)P(x,t)-[g_{eff}(x)\partial_{x}g_{eff}(x)P(x,t)], 
\end{equation}
where,
\begin{equation} \label{geff}
g_{eff}(x)= \sqrt{K_{B}T_{1}g_{1}^{2}(x)+K_{B}T_{2}g_{2}^{2}(x) },
\end{equation}
and $H(x)$ can be related to an effective potential,
\begin{equation} \label{Veff}
V_{eff}(x)= -\int^{x}H(x')dx'.
\end{equation}

Then we have to solve the equation (\ref{flux}) in the steady state for a constant flux $J$ 
with periodic conditions. The first step is to reduce this equation to a 
Bernoulli form which can be formally integrated. By imposing periodic 
boundary conditions, $P(x)=P(x+L)$ (where $L=\frac{2\pi}{d}$), we get,
\begin{equation} 
\label{JP}
P_{0}(1-e^{\beta(L)})= J\int_{0}^{L} dx \frac{e^{-\beta(x)}}{g_{eff}(x)},
\end{equation}
where $P_{0}$ is a constant that can be found by using the normalization condition
$\int_{0}^{L}P(x)dx=1$, and $\beta$ is a relevant function whose
expression is, 
\begin{equation} 
\label{beta1}
\beta(x)=\int_{0}^{x}dx'\frac{-H(x')}{g_{1}^{2}(x')+(T_{2}/T_{1})g_{2}^{2}(x')},
\end{equation}
where we take $K_{B}T_{1}=1$. Then the mean velocity $v \doteq \langle \dot{x} \rangle \simeq \langle
\dot{\theta_{1}} \rangle $ (since
$ \langle \dot{\theta_{1}} \rangle \simeq \langle \dot{\theta_{2}} \rangle $) is found to be proportional to
$(1-e^{\beta(L)})$. What is left to do is to find the $\beta$ integral for every $\phi$.

To simplify the calculation of the integral, we now make an expansion in powers of
$\frac{1}{k}$. Since the value of $k$ that makes the motor run faster is around
$k=100$, one can safely suppose that the terms of the order $(\frac{1}{k})^{2}$ and so
on will not notably contribute to the integral for $d<5$. Then, the integral
(\ref{beta1}) turns out to be (\ref{beta}), which is evaluated numerically.

\end{document}